\documentclass[nojss]{jss}

\usepackage{orcidlink,thumbpdf,lmodern}

\usepackage{framed}
\usepackage{longtable}
\usepackage{booktabs}

\usepackage{tabularx}
\usepackage{array}

\author{Hugo Morvan\\ Link\"oping University
   \And Jonas Agholme \orcidlink{0009-0000-2041-943X}\\Link\"oping University
   \And Bj\"orn Eliasson\\G\"oteborg University
   \And Katarina Olofsson \\Region V\"astra Götaland
   \AND Ludger Grote \\G\"oteborg University\\
    \And Fredrik Iredahl\textsuperscript{\dag} \orcidlink{0000-0002-4245-7565}\\Link\"oping University
   \And Oleg Sysoev\textsuperscript{\dag *} \orcidlink{0000-0002-3092-4162}\\Link\"oping University}
   
\Plainauthor{Hugo Morvan, Jonas Agholme, Bj\"orn Eliasson, Katarina Olofsson,  Ludger Grote, Fredrik Iredahl, Oleg Sysoev}

\title{\pkg{bigMICE}: Multiple Imputation of Big Data}

\Plaintitle{bigMICE: Multiple Imputation of Big Data}
\Shorttitle{\pkg{bigMICE}: Multiple Imputation of Big Data}

\Abstract{
    Missing data is a prevalent issue in many applications, including large medical registries such as the Swedish Healthcare Quality Registries, potentially leading to biased or inefficient analyses if not handled properly.  Multiple Imputation by Chained Equations (MICE) is a popular and versatile method for handling multivariate missing data but traditional implementations face significant challenges when applied to big data sets due to computational time and memory limitations.
    
    To address this, the \pkg{bigMICE} package was developed, adapting the MICE framework to big data using Apache \pkg{Spark MLLib} and \pkg{Spark ML}. Our implementation allows for controlling the maximum memory usage during the execution, enabling processing of very large data sets on a hardware with a limited memory, such as ordinary laptops. 
    
    The developed package was tested on a large Swedish medical registry to measure memory usage, runtime and dependence of the imputation quality on sample size and on missingness proportion in the data. In conclusion, our method is generally more memory efficient and faster on large data sets compared to a commonly used MICE implementation. We also demonstrate that working with very large datasets can result in high quality imputations even when a variable has a large proportion of missing data. This paper also provides guidelines and recommendations on how to install and use our open source package.

}

\Plainkeywords{missing data, multiple imputation, MICE, big data, Apache Spark, R}

\begin{document}

\renewcommand{\thefootnote}{\fnsymbol{footnote}}
\footnotetext[2]{ : shared last authorship}
\footnotetext[1]{ : corresponding author, email: oleg.sysoev@liu.se}


\section[Introduction]{Introduction}

Missing data is a prevalent issue in many applications that can significantly impede data analysis across various scientific and engineering domains \cite{Yang_2020}. In research, the absence of values for certain attributes within a data set is a common occurrence \cite{Wang_2020, Austin_2020,Arefin_2024}, arising from a multitude of reasons, including clerical errors during data entry, system malfunctions, participant non-disclosure, participants dropouts from studies, data corruption, or variations in data collection procedures \cite{Austin_2020, Cesare_2022,Arefin_2024}. The presence of missing data poses substantial challenges, as many algorithms require a complete matrix of indicators without gaps \cite{Wang_2020}, and statistical results derived from incomplete data sets, particularly those with nonrandom missing values, can be biased \cite{Hasan_2021}.

\subsection{Problem Statement}

Many big data sets in society, including medical and administrative registries, are invaluable resources for research and decision-making. However, these data sets often contain missing values.
Traditional approaches to handling missing data have proven inadequate for big data contexts. Typical approaches for handling missing data include case deletion (also known as complete case analysis or list-wise deletion) and simple imputation methods such as replacing missing values with the mean, median or mode \cite{Pampaka_2016, Grund_2016, Austin_2020}. However, these single imputation methods do not account for the uncertainty associated with missing data and can attenuate variances and covariances, potentially leading to artificially narrow confidence intervals and inflated Type I error rates \cite{Austin_2020}.

Multiple Imputation by Chained Equations (MICE) is a popular and flexible method for imputing missing values \cite{Buuren_2007}. It works by generating multiple plausible values for each missing data point, creating multiple complete data sets for analysis. This approach ensures that the uncertainty of the predictions is properly taken into account, leading to more valid and more reliable downstream results.

Despite the widespread availability of MICE implementations in various programming languages and statistical software packages, a critical limitation emerges when applying these methods to large-scale data sets.
The application of MICE to big data sets can lead to huge computational times due to the sheer size of these data sets (sometimes hundreds of gigabytes of data) and the iterative nature of MICE. However, the main problem is that computations may fail because the amount of memory required exceeds the Random Access Memory (RAM) space of a given computational resource. As modern society is constantly collecting new data,  data registries are getting bigger every year, and it becomes of a critical importance to be able to create accurate imputations of these large data.

\subsection{Existing Software Limitations}

There are many MICE implementations in various programming languages and statistical software packages.

The R statistical computing environment offers one of the most comprehensive collection of MICE implementations, including the \textbf{mice}  package \cite{miceR} which commonly serves as a reference implementation for the MICE algorithm. Other notable \proglang{R} implementations include \textbf{parlMICE} for parallel computation, \textbf{miceRanger} using random forests, \textbf{Amelia} using bootstrapping and the EM algorithm \cite{Pampaka_2016, Khan_2020, Little_2019, Lo_2019}, and specialized packages like \textbf{mitml} for multilevel data \cite{Grund_2016}.

Beyond R, implementations exist in Python (\textbf{scikit-learn}'s IterativeImputer, \textbf{miceforest}), SAS (\textbf{PROC MI}), Stata (\textbf{ice} command), and various other platforms \cite{He_2021, Kombo_2017, Pampaka_2016, miceR}. Recognizing the computational demands of MICE, efforts have been made to improve computational time through parallelization. While parallelization of the imputation process itself is not possible due to its iterative nature, it is possible to parallelize the different imputations as they are independent, which is the approach taken in the \pkg{parlMICE} package. However, this approach still does not provide a solution to the fundamental memory challenges when big data are used.

A MICE implementation capable of handling large data sets (in particular, large medical records) in the presence of limited computational resources has not yet been developed.

\subsection[The bigMICE solution]{The \pkg{bigMICE} solution}

To address the gap in existing methodologies related to computing multiple imputations on large data sets, we have developed our package \pkg{bigMICE}. This package provides a framework for distributed missing data imputation, leveraging Apache \pkg{Spark}'s capabilities to fit computations to a given RAM space by dynamically exchanging information between RAM and the hard drive \cite{apache_spark}. This makes it possible to handle large data sets with millions of observations on an ordinary computer having typical technical characteristics, for example, 16Gb of RAM or even less. 

To handle large datasets, the package \pkg{bigMICE} integrates the MICE framework with the Apache \pkg{Spark} and \pkg{MLLib} functionalities. Existing methodologies lack the capability to perform multiple imputation on large data sets natively within \pkg{Spark}'s distributed computing environment. The package incorporates several key design principles: distributed data handling using \pkg{Spark} dataframes, scalable regression models leveraging \pkg{Spark MLLib}'s optimized implementations, memory-efficient storage strategies, and iterative processing adapted to work within \pkg{Spark}'s directed acyclic graph (DAG) execution model.

\subsection{Contributions and Scope}

This paper introduces a software implementation of the MICE framework designed specifically for big data sets. Our package \pkg{bigMICE} addresses a critical gap in the statistical software ecosystem by enabling multiple imputation of data sets that exceed the memory capacity of traditional implementations by employing the Apache \pkg{Spark} environment.

The primary contributions of this work include:
\begin{enumerate}
    \item \textbf{Novel Software Architecture}: A re-implementation of the MICE algorithm within the \pkg{Spark} distributed computing framework, maintaining statistical validity while achieving scalability.
    
    \item \textbf{Memory Management Solutions}: Innovative approaches to handle the memory demands of multiple imputation in big data contexts, including efficient storage strategies and parameter-based pooling methods.
    
    \item \textbf{Integration with Existing Ecosystems}: Seamless integration with the \proglang{R} statistical computing environment through the \proglang{R} interface to \pkg{Spark} (the package \pkg{sparklyr}). 
    
    \item \textbf{Comprehensive Evaluation Framework}: Systematic assessment of computational performance, memory efficiency, and statistical validity for various data set sizes and complexity levels.
    
    \item \textbf{Open Source Implementation and Release}: A fully documented open-source package that can be extended and adapted for various big data imputation scenarios.
\end{enumerate}

\subsection{Paper Organization}

The remainder of this paper is organized as follows.
Section 2 provides background information on the MICE methodology and \pkg{Spark} distributed computing frameworks. 
Section 3 provides guidelines on how to install and use the package. 
Section 4 provides a description of the software architecture and details of the implementation of the \pkg{bigMICE} package.
Section 5 contains results of comprehensive computational experiments evaluating the performance, scalability, and statistical properties of the implementation using a real-world medical registry data set. 
Section 6 discusses the results of the experiments and provides directions for future development and extensions of the framework.

The software implementation, documentation, and reproducible examples are available through \href{https://github.com/bigcausallab/bigMICE}{github.com/bigcausallab/bigMICE}.

\section[Background]{Background}

\subsection{Multiple Imputation by Chained Equations: Theoretical Foundation}

\subsubsection{Data missingness}

To understand the nature of missing data, it is crucial to consider the underlying mechanisms that cause the missingness. Missing values can be classified into three different types, depending on whether the fact that a variable is missing is related to the underlying values of the variables in the data set \cite{Little_2019, White_2011}:

\begin{itemize}
    \item \textbf{Missing Completely At Random (MCAR)}: the probability of data being missing does not depend on observed or unobserved data. MCAR implies that the probability of missing data is independent of both the observed and unobserved variables \cite{Hasan_2021,Wang_2020,Pampaka_2016,Grund_2016,Khan_2020,Austin_2020}
    \item \textbf{Missing At Random (MAR)}: The probability of data being missing does not depend on the unobserved data, conditional on the observed data. Data are considered MAR if, after accounting for all observed variables, the probability of a variable being missing is independent of unobserved data \cite{Hasan_2021,Wang_2020,Grund_2016,Khan_2020,Austin_2020,miceR}. In other words, the missingness can be predicted by other observed variables in the data set \cite{Wang_2020}.
    \item \textbf{Missing Not At Random (MNAR)}: The probability of missing data depends on the values of the missing variables themselves, even after conditioning on the observed data \cite{Hasan_2021,Wang_2020,Grund_2016,Khan_2020,Austin_2020,miceR}. Analyses of data under MNAR condition are particularly challenging, as they require modeling the missing data mechanism itself, which is often unknown. \cite{Wang_2020}.
\end{itemize}

It is not possible to distinguish between MAR and MNAR from the observed data alone, although the MAR assumption can be made more plausible by collecting additional variables and including them in the analysis \cite{White_2011}.

\subsubsection{Multiple Imputation Framework}

Multiple Imputation (MI) is a general statistical method designed for the analysis of incomplete data sets \cite{Buuren_2007,Pampaka_2016,Kombo_2017,Austin_2020}.  MI offers a strategy to provide valid inferences for statistical estimates in missing data entries \cite{Buuren_2007}. The main principles of MI were proposed by Rubin \cite{Rubin_1987}, establishing it as a leading approach to handling missing data. The core idea of MI is to use the distribution of the observed data to estimate a set of plausible values for the missing data, incorporating random components to reflect their uncertainty and creating multiple ($m$) complete data sets. These data sets are then individually analyzed and the results are combined to obtain overall estimates and variance-covariance matrices using Rubin's rules \cite{White_2011}.

A statistical analysis using MI typically involves three major steps \cite{Buuren_2007,Pampaka_2016}:

\begin{itemize}
    \item \textbf{Imputation}: The first step involves specifying an imputation model and generating plausible synthetic data values, called imputations, for missing values in the data \cite{Buuren_2007,Pampaka_2016,miceR}. This process results in a number $m$ of complete data sets where missing data are replaced by random draws from a distribution of plausible values \cite{Buuren_2007,Pampaka_2016}. The number of imputations, $m$, typically ranges from 3 to 10 \cite{Buuren_2007}, although more may be needed for certain analyses \cite{Grund_2016}.

    \item \textbf{Analysis}: The second step consists of analyzing each of the $m$ imputed data sets using a statistical method that will estimate the quantities of scientific interest \cite{Buuren_2007,Pampaka_2016}, for example, the coefficients of a regression model. This results in $m$ analyzes, which differ only due to variations in the imputations \cite{Buuren_2007}. MI allows the analyst to use techniques designed for complete data \cite{Pampaka_2016}.

    \item \textbf{Pooling}: The third step involves pooling the $m$ estimates into one single estimate \cite{Buuren_2007}. This process combines the variation within and across the $m$ imputed data sets \cite{Buuren_2007,Kombo_2017}. Under fairly liberal conditions, this step produces statistically valid estimates that reflect the uncertainty caused by missing data in the width of the confidence interval \cite{Buuren_2007,Pampaka_2016}.
\end{itemize}

The pooling step follows Rubin's rules, which allow combining the estimated parameters in the $m$ imputations into an overall estimate and a variance-covariance matrix that incorporates both the within-imputation variability (uncertainty about the results from one data set) and between-imputation variability (reflecting the uncertainty due to the missing information) \cite{White_2011}.

\subsubsection{Multiple Imputation by Chained Equations}

Within the framework of multiple imputation, two main approaches have emerged for imputing multivariate data: Joint Modeling (JM) and Multiple Imputation by Chained Equations (MICE) \cite{Buuren_2007,miceR}. While JM is based on specifying a multivariate distribution for missing data \cite{miceR}, MICE, also known as Fully Conditional Specification (FCS)  or sequential regression multivariate imputation \cite{Buuren_2007,miceR,Yang_2020,Austin_2020}, takes a different route.
The MICE algorithm aims to overcome the limitations of univariate imputation by using a set of conditional models, one for each variable with missing data. 

Some notations are introduced first. Let $Y_j$ be $j^{th}$ column (variable) of dataset $Y$ with $p$ variables, and $Y^{mis}_j$ is a part of $Y_j$ with missing values,  and $Y^{obs}_j$ is a part of $Y_j$ with the observed (non-missing) values. Denote $Y^{obs}=\{Y^{obs}_1, \ldots, Y^{obs}_p \}$ and $Y^{mis}=\{Y^{mis}_j, \ldots\, Y^{mis}_p\}$. MICE algorithm acts by imputing missing values in every variable multiple times (multiple iterations). Denote $Y^t_j$ variable $Y_j$ imputed at iteration $t$ and $Y_{-j}^t = (Y_{1}^{t},\ldots,Y_{j-1}^{t},Y_{j+1}^{t-1},\ldots,Y_{p}^{t-1})$.

Initially, missing values are filled in by simple random sampling with replacement of the observed values. These initial random draws are iteratively replaced as the MICE procedure progresses.
In the first iteration, the first variable with missing values, $Y_{1}$, is regressed on all other variables $Y_{2}, \ldots, Y_{k}$, using only cases with observed $Y_{1}$, and the missing values are replaced by simulated draws from the posterior predictive distribution of $Y_{1}$. The process continues sequentially for each variable with missing data, such as $Y_{2}$, which is regressed on $Y_{1}, Y_{3}, \ldots, Y_{k}$, using the imputed $Y_{1}$, and so forth. This process is repeated a number of times or ideally until convergence to produce a complete imputed data set, and then the entire procedure is repeated $m$ times to generate $m$ multiply imputed data sets.

More generally, for each incomplete variable, a conditional model needs to be specified as:
$$P(Y_{j} \mid Y^{obs}_j, Y_{-j}, \theta_{j}),$$

where $Y_{j}$ is the variable that is imputed,  $Y_{-j}$ denotes the remaining variables variables in the data set, and $\theta_{j}$ are the parameters of the conditional model \cite{Buuren_2007}. 

Starting from initial guesses, MICE imputation proceeds by iterating over all specified conditional models \cite{Buuren_2007}. If the joint distribution implied by these conditional distributions exists, the algorithm functions as a Gibbs sampler \cite{miceR,Buuren_2007}.

\begin{figure}[ht]
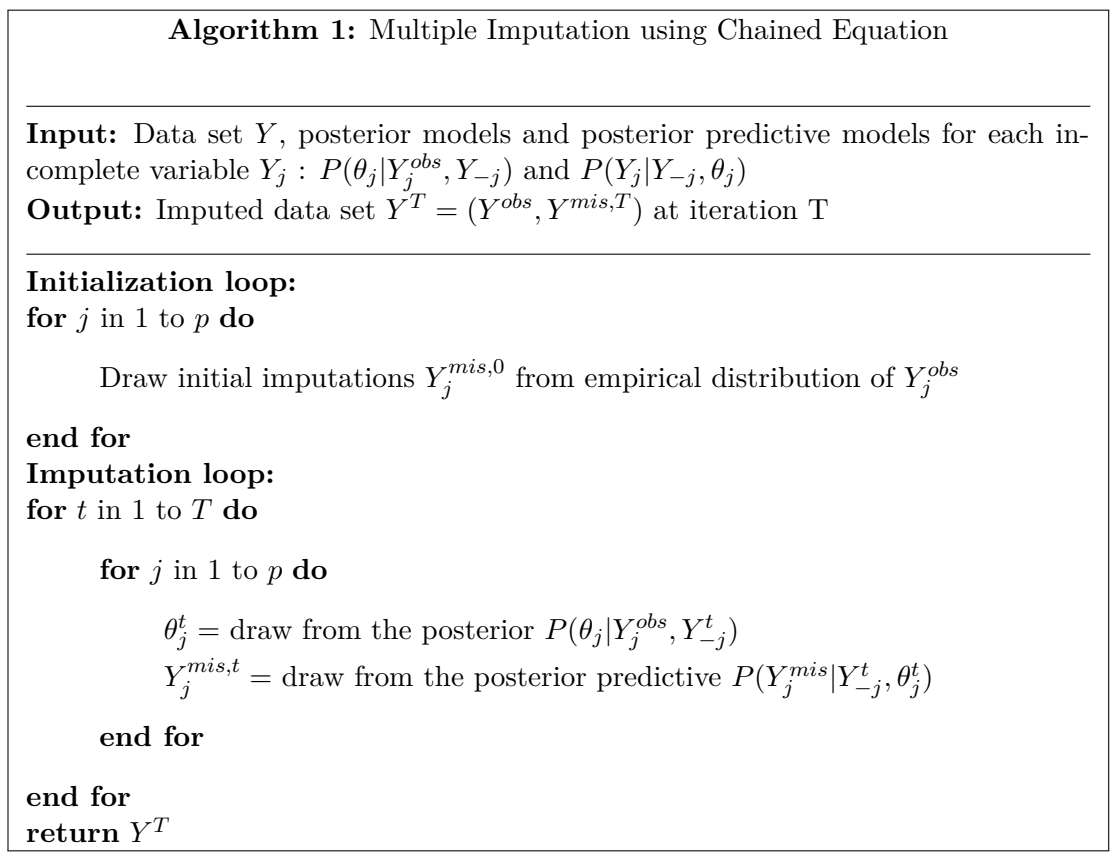

    \centering
    \fbox{
        \begin{minipage}{0.9\textwidth}
            \begin{center}
                \textbf{Algorithm 1:} Multiple Imputation using Chained Equation
            \end{center}
            \rule{\linewidth}{0.4pt}
            \textbf{Input:} Data set $Y$, 
             posterior models and posterior predictive models for each incomplete variable $Y_j$ : $P(\theta_j | Y_{j}^{obs}, Y_{-j})$ and $P(Y_j | Y_{-j}, \theta_j)$ \\
            \textbf{Output:} Imputed data set $Y^T = (Y^{obs},Y^{mis,T})$ at iteration T
            
            \rule{\linewidth}{0.4pt}

            \textbf{Initialization loop:}\\
            \textbf{for} $j$ in 1 to $p$ \textbf{do}
            \begin{quote}
                Draw initial imputations $Y_j^{mis,0} $ from empirical distribution of $Y_j^{obs}$
            \end{quote}
            \textbf{end for}

            \textbf{Imputation loop:}\\
            \textbf{for} $t$ in 1 to $T$ \textbf{do}
            \begin{quote}
                \textbf{for} $j$ in 1 to $p$ \textbf{do}
                \begin{quote}
                    $\theta_j^t = $ draw from the posterior $P(\theta_j | Y_{j}^{obs}, Y_{-j}^t)$
                    
                    $Y_j^{mis,t} = $ draw from the posterior predictive $P(Y_j^{mis} | Y_{-j}^{t}, \theta_j^t)$
                \end{quote}
                \textbf{end for}
            \end{quote}
            \textbf{end for}
            
            \textbf{return} $Y^T$    
        \end{minipage}
    }
    \caption{Multiple Imputation using Chained Equation (MICE)}
    \label{alg:mice}
\end{figure}

MICE offers substantial flexibility for creating multivariate models and is often more feasible to implement than JM, especially in the presence of complex data structures such as skip patterns, restrictions, and bounds, which are common in large-scale survey data \cite{Buuren_2007,miceR,He_2021}. It enables the use of variable-specific imputation methods, such as linear regression for continuous data, logistic regression for binary data, or proportional odds regression for ordinal data \cite{Buuren_2007,RescheRigon_2018,Kombo_2017,He_2021}, and supports the incorporation of constraints to maintain logical consistency in imputed values \cite{Buuren_2007,Buuren_2006}. The procedure for generating one imputation using the MICE method is described in Algorithm \ref{alg:mice}. 

The "chained equations" in MICE refer to the regression models that are used to iteratively impute missing values in each of the variables. Using multiple regression models allows for handling multiple types of variable types. Common specifications of conditional models include logistic regression for binary variables, linear regression or random forest regression for continuous variables, and multinomial logistic regression or random forest classification for categorical variables.

\subsection[Apache Spark and Distributed Computing Solutions]{Apache \pkg{Spark} and Distributed Computing Solutions}

The immense volume of data generated has led to the creation of specialized big data platforms designed to manage and analyze this information. Big data platforms are comprehensive frameworks that enable organizations to store, process, and analyze both structured and unstructured data, by leveraging distributed computing, parallel processing, and advanced analytics. Big data platforms offer a variety of features, including robust storage solutions and distributed processing capabilities to handle large data volumes across multiple nodes. They ensure fault tolerance for continuous data availability and uninterrupted processing, even during failures.

\subsubsection[Apache Spark framework]{Apache \pkg{Spark} Framework}

\pkg{Spark} \cite{apache_spark} is a powerful open source platform designed for large-scale data processing, particularly well suited for iterative statistical and machine learning tasks \cite{Meng_MLLib, Zaharia_2016_spark}. \pkg{Spark} improves on the traditional MapReduce paradigm \cite{dean2008mapreduce} by offering better data management, leading to significantly faster performance \cite{Salloum2016_spark}. At its core, \pkg{Spark} uses resilient distributed data sets (RDDs) as a fundamental abstraction for efficient data sharing between computations \cite{Zaharia_2016_spark, Salloum2016_spark}. With \pkg{Spark}, data can be processed through a more general directed acyclic graph (DAG) of operators using rich sets of transformations and actions. It automatically distributes the data across the cluster and parallelizes the required operations. It supports a variety of transformations that make data pre-processing easier, especially when it is becoming more difficult to examine big data sets \cite{Salloum2016_spark}.

In \pkg{Spark}, efficient data reuse and fault tolerance are supported through mechanisms such as caching, persisting, and check-pointing. Caching is used to temporarily store data in the memory, which is particularly useful when the same data set is accessed repeatedly across multiple operations, and speeds up processing significantly. Persisting serves a similar role but allows more flexibility, enabling the user to decide whether the data should be stored in memory, on disk, or in a combination of both, depending on resource constraints. Check-pointing, on the other hand, is a more robust process designed for long and complex computation chains. It saves data to a reliable storage system, such as the Hadoop Distributed File System (HDFS), which helps prevent data loss and reduces computational overhead by truncating the lineage of transformations.

Checkpointing is the major asset from \pkg{Spark} that enables the computation of large operations on large data sets on computers with small RAM capacity. It is widely used in \pkg{bigMICE} to avoid memory problems, such as stack overflow errors.

\subsubsection[Spark MLLib for Machine Learning]{\pkg{Spark MLLib} for Machine Learning}

A key component of \pkg{Spark} is \pkg{MLLib}, its open-source distributed machine learning library. \pkg{MLLib} provides efficient functionality for a wide range of learning settings and includes several underlying statistical, optimization, and linear algebra primitives \cite{Meng_MLLib}. One of the main advantages of using \pkg{MLLib} is that it includes many regression and classification models, and algorithms that are optimized to run efficiently on the \pkg{Spark} platform, including linear regression, logistic regression, decision trees, and random forests, all of which are commonly used imputation models in the MICE framework.

The iterative nature of many machine learning algorithms, such as logistic regression involving procedures such as gradient descent to optimize an objective function, makes \pkg{Spark}'s ability to perform in-memory computations on distributed data sets crucial for efficiency \cite{Zaharia_2016_spark, Zaharia_2012_RDD}. Because \pkg{Spark} can cache the data in RAM across these iterations, it avoids the overhead of reloading data from disk in each step, leading to significant speedups compared to systems like Hadoop's MapReduce, especially for iterative algorithms \cite{Zaharia_2016_spark, Zaharia_2012_RDD}. \pkg{Spark}'s \pkg{MLLib} library provides optimized implementations of these algorithms, further simplifying their use in big data machine learning pipelines.


\subsubsection[The sparklyr interface]{The \pkg{sparklyr} interface}

The package \pkg{sparklyr} is an \proglang{R} interface for Apache \pkg{Spark}, enabling data scientists and analysts to leverage distributed computing capabilities of \pkg{Spark} directly from the \proglang{R}  environment. It provides tools for connecting to \pkg{Spark} clusters, manipulating large-scale data sets using \pkg{dplyr}-style syntax, performing machine learning through \pkg{Spark MLLib}, and executing custom \proglang{R} or \proglang{SQL} code. By seamlessly integrating with both \proglang{R}  and \pkg{Spark}, \pkg{sparklyr} facilitates scalable data processing and modeling, making it well-suited for research and applications involving big data analytics without requiring researchers to leave the familiar \proglang{R}  environment.

\subsection[Methodological Adaptations for bigMICE]{Methodological Adaptations for \pkg{bigMICE}}

The implementation of \pkg{bigMICE} incorporates several key modifications to the traditional MICE workflow:
\begin{itemize}
    \item \textbf{Initialization Methods}: Two approaches were developed for initializing missing values: summary statistics (mean, mode, median) and random sampling from observed values, both implemented using \pkg{Spark} operations to allow scalable computation.
    
    \item \textbf{Classification and Regression Model Integration}: The package \pkg{bigMICE} employs specialized functions that utilize various regression models available in \pkg{Spark MLLib}, including linear regression for continuous variables, logistic regression for binary variables, multinomial logistic regression for categorical variables, and random forest methods for both regression and classification tasks.
    
    \item \textbf{Rubin's Rules Implementation}: To manage memory demands, the package applies Rubin's rules by storing only model parameters for pooling results, rather than maintaining entire imputed data sets in memory throughout the process.
    
    \item \textbf{Checkpointing}: To prevent execution failures due to occasional memory bottlenecks, we apply  \pkg{Spark}'s checkpointing.

    \item \textbf{Customization of the running environment}: Through the \pkg{Spark} connection configuration, users of \pkg{bigMICE} can specify the resources (for example, number of CPU cores and memory limit) allocated to the computation. This provides a great flexibility across different execution environments, from distributed clusters to small laptops, making it possible to use the \pkg{bigMICE} package regardless of the available RAM or CPU, provided sufficiently large hard drive space.
\end{itemize}

\section[Setup and Guidelines]{Setup and Guidelines}

Several other pieces of software must be installed before using \pkg{bigMICE}. The following section details the steps required and shows a simple example of the package usage.

\subsection{Setup}

\subsubsection[sparklyr package]{\pkg{sparklyr} package}

The package \pkg{bigMICE} integrates with the \pkg{sparklyr} framework, allowing users to manage \pkg{Spark} clusters or standalone machines with multiple cores directly from \texttt{R}. The package was tested with \pkg{Spark} version 4.0.0 and \pkg{sparklyr} version 1.9.1, although earlier versions of \pkg{Spark} (< 4.0.0) may be required for compatibility with previous releases of \pkg{sparklyr}. 

\pkg{Spark} can be installed directly from \proglang{R} by first installing \pkg{sparklyr}, as follows: 

\begin{CodeChunk}
\begin{CodeInput}
R> install.packages("sparklyr") # version 1.9.1
R> options(timeout = 6000) # for slow internet connections
R> library("sparklyr")
R> spark_install(version="4.0.0") # compatible with sparklyr 1.9.1
\end{CodeInput}
\end{CodeChunk}

Once both \pkg{sparklyr} and \pkg{Spark} are installed, users can connect to \pkg{Spark} while specifying the desired resources to be used:

\begin{CodeChunk}
    \begin{CodeInput}
R> conf <- spark_config()
R> conf$`sparklyr.shell.driver-memory` <- "10G" # How much memory
R> conf$`sparklyr.cores.local` <- 4 # How many CPU cores
R> sc = spark_connect(master = "local", config = conf)
    \end{CodeInput}
\end{CodeChunk}

\subsubsection{Hadoop Distributed File System}

In addition to \pkg{Spark}, it is recommended to install Hadoop to facilitate the use of the Hadoop Distributed File System (HDFS). HDFS integration enables checkpointing, which improves computational stability and resilience. If HDFS is not available, users can still use the package by disabling checkpointing using the parameter \texttt{checkpointing = FALSE}.


Given an HDFS directory, the checkpointing directory can be set up as follows after connecting to your \pkg{Spark} Session:

\begin{CodeChunk}
    \begin{CodeInput}
R> spark_set_checkpoint_dir(sc, "/path/to/your/HDFSdirectory/" )     
    \end{CodeInput}
\end{CodeChunk}

\subsubsection{Development version of bigMICE (GitHub)}

The \pkg{bigMICE} package is hosted on GitHub and can be installed via the \pkg{devtools} package: 

\begin{CodeChunk}
    \begin{CodeInput}
R> install.packages("devtools")
R> library("devtools")
R> devtools::install_github("bigcausallab/bigMICE")
    \end{CodeInput}
\end{CodeChunk}

\subsection{Example Usage}

The following example demonstrates a typical workflow using the \pkg{bigMICE} package when executed on a single machine with spark installed. This includes setting up the environment, loading data, defining variable types and an analysis model, and performing multiple imputation.

\paragraph{Loading Required Libraries:}

In this example we load \pkg{bigMICE}, \pkg{sparklyr}  and \pkg{dplyr}, to facilitate data manipulation.
\begin{CodeChunk}
\begin{CodeInput}
R> library("bigMICE")
R> library("sparklyr")
R> library("dplyr")
\end{CodeInput}
\end{CodeChunk}

\paragraph{Initializing a Local Spark Session:}

We specify a \pkg{Spark} configuration for a local session. We set the \pkg{Spark} memory limit to 10Gb of memory and 4 CPU cores:
\begin{CodeChunk}
    \begin{CodeInput}
R> conf <- spark_config()
R> conf$`sparklyr.shell.driver-memory` <- "10G"
R> conf$spark.memory.fraction <- 0.8
R> conf$`sparklyr.cores.local` <- 4
R> sc <- spark_connect(master = "local", config = conf)
    \end{CodeInput}
\end{CodeChunk}

\paragraph{Loading and Preparing the data set:}

For this example we use the "boys" dataset from \pkg{mice} package, which can be downloaded at \url{https://github.com/amices/mice/blob/master/data/boys.rda}. We first start by converting this dataset to a \texttt{csv} format and save it to our working directory.
\begin{CodeChunk}
    \begin{CodeInput}
R> data <- load("boys.rda")
R> write.csv(boys, "data.csv", row.names = FALSE)
    \end{CodeInput}
\end{CodeChunk}
Assuming the data set has now been saved in your working directory as \texttt{data.csv}, it is loaded into \pkg{Spark} and certain variables are excluded from the analysis.

\begin{CodeChunk}
\begin{CodeInput}
R> sdf <- spark_read_csv(sc, "data", "data.csv", 
+     header = TRUE, infer_schema = TRUE, null_value = "NA") %>%
+     select(-all_of(c("hgt", "wgt", "bmi", "hc"))) # removed variables
\end{CodeInput}
\end{CodeChunk}

\paragraph{Defining Variable Types and Analysis Formula:}

Variable types must be explicitly specified, and a formula is defined for the downstream analysis.

\begin{CodeChunk}
\begin{CodeInput}
R> variable_types <- c(
+     age = "Continuous_float", 
+     gen = "Nominal", 
+     phb = "Nominal",
+     tv  = "Continuous_int",
+     reg = "Nominal"
+ )
R> analysis_formula <- as.formula("phb ~ age + gen + tv + reg")
\end{CodeInput}
\end{CodeChunk}

\paragraph{Performing Multiple Imputation:}

The imputation is carried out using \texttt{mice.spark} with three imputed data sets (\texttt{m = 3}) and 2 iterations (\texttt{maxit = 2}) per imputation. Checkpointing is disabled in this case:

\begin{CodeChunk}
\begin{CodeInput}
R> imputation_results <- bigMICE::mice.spark(
+     data = sdf,
+     sc = sc,
+     variable_types = variable_types,
+     analysis_formula = analysis_formula,
+     m = 3,
+     maxit = 2,
+     checkpointing = FALSE
+ )
\end{CodeInput}
\end{CodeChunk}

\paragraph{Extracting Results}

The following elements are returned from the \texttt{mice.spark} function.

The pooled results of the analysis:

\begin{CodeChunk}
    \begin{CodeInput}
R> print(imputation_results)
    \end{CodeInput}
    \begin{CodeOutput}
Multiple Imputation Results
==========================

Number of imputations: 3 
Total imputation time: 299.94 seconds
Average time per imputation: 99.98 seconds

Pooled Parameter Estimates (Rubin's Rules)
==========================================
            Estimate Within_Var Between_Var Total_Var      SE t_stat
(Intercept)        0    97.2486   2625.7130 3598.1993 59.9850      0
age                0     0.3800     10.2599   14.0599  3.7497      0
tv                 0     0.0003      0.0073    0.0100  0.1001      0

Diagnostic Information:
-----------------------
(Intercept)    : r=36.000, lambda=0.973, df=2.1
age            : r=36.000, lambda=0.973, df=2.1
tv             : r=36.000, lambda=0.973, df=2.1

Notes:
------
- SE: Standard Error (sqrt of Total_Var)
- t_stat: t-statistic for testing parameter = 0
- r: Relative increase in variance due to nonresponse
- lambda: Fraction of missing information
- df: Degrees of freedom for t-distribution
- Use show_individual=TRUE to see results from each imputation
    \end{CodeOutput}
\end{CodeChunk}

The model parameters estimated for each imputed data set:
\begin{CodeChunk}
    \begin{CodeInput}
imputation_results$model_params
    \end{CodeInput}
\end{CodeChunk}

And some statistics on the imputation process, such as imputation time:
\begin{CodeChunk}
    \begin{CodeInput}
imputation_results$imputation_stats
    \end{CodeInput}
\end{CodeChunk}

Additionally, it is possible to return the complete imputed data sets by using the \texttt{mice.spark.plus} function instead of \texttt{mice.spark} function (at a greater memory cost), followed by:  
\begin{CodeChunk}
    \begin{CodeInput}
imputation_results$imputations
    \end{CodeInput}
\end{CodeChunk}

\section{Package Description}

Although there are various MICE implementations, there is a lack of dedicated tools capable of efficiently handling the scale and the distributed nature of data managed by \pkg{Spark} dataframes. To address this need, the \pkg{bigMICE} package was conceived and developed. The package provides a framework for distributed missing data imputation, taking advantage of \pkg{Spark}'s capabilities. A number of modifications were done to the standard MICE framework within \pkg{bigMICE} to enable seamless integration with \pkg{Spark} dataframes and the \pkg{Spark MLLib} library. These modifications are described in this section.

\subsection{Initialization method}

The original MICE framework enables different approaches for the initial imputation of missing values, including both random sampling of observed values and imputation using summary statistics such as mean, mode, or median to initialize the missing values. 
Consequently, two methods were developed in \pkg{bigMICE} to initialize the missing values with summary statistics or random sampling of the observed values.

The \texttt{impute\_with\_MeMoMe} function can be used for the initialization of missing values using summary statistics. This function takes a \pkg{Spark} connection, a \pkg{Spark} DataFrame, and a named vector that specifies the imputation method (mean, mode, median, or none) for each column. The selection of the appropriate initial imputation method for each variable is guided by its variable type, provided by the user.
A predefined dictionary within the \texttt{mice.spark} function maps different types of variable to their corresponding initial imputation methods. For instance, binary, nominal, and ordinal variables are initialized using the mode, while integer and count variables use the median, and float variables are initialized with the mean. Variables designated with types such as code, date/time, and string are set to "none", indicating that no initial imputation is performed for these variables as summary statistics of those variable types are not possible. This mapping is then used to create the initialization method vector, which specifies the initial imputation method for each column in the input \pkg{Spark} DataFrame.
The \texttt{impute\_with\_MeMoMe} function then proceeds to calculate the respective mean, mode, or median from the observed values for each specified column. These calculated values are subsequently used to replace the missing values in the corresponding columns of the \pkg{Spark} DataFrame, resulting in an initially imputed DataFrame. 

The \texttt{impute\_with\_random\_samples} function can be used for the initialization of missing values using random samples with replacement.  This function takes f the \pkg{Spark} connection and a \pkg{Spark} DataFrame as arguments. The function utilizes \pkg{sparklyr}'s \texttt{sdf\_sample} function to perform sampling with replacement on the observed values. We do some extra steps to circumvent a flaw of the function. \texttt{sdf\_sample} does not take for input a precise number of desired sample, but rather a fraction of the input vector and thus sometimes lacks precision and returns fewer samples than the specified fraction. Hence, to use this function, the fraction needed is calculated by dividing the number of missing values by the number of observed values. Then, the function is called to sample with replacement from the observed values to obtain the desired amount to replace the missing values. If the correct number of values are sampled, they are used to replace the missing values in the data. If not, the sampler function is recalled until the correct amount is sampled. Although it may require multiple sampling attempts to obtain the desired number of samples, it is in practice still more efficient than using other (more accurate) sampling functions that are not optimized for \pkg{Spark} data and thus much slower.

This initial imputation step can either be executed by the \texttt{mice.spark} function itself or can be provided by the user as input (using the \texttt{imp\_init} optional parameter), offering more flexibility. By default, \texttt{mice.spark} uses \texttt{impute\_with\_random\_samples} to initialize the missing values. This initialized dataframe serves as the starting point for the subsequent iterative imputation process within the MICE algorithm.

\subsection{Imputation methods}

The developed package employs a suite of specialized functions to perform individual imputation steps using various regression and classification models available in \pkg{Spark MLLib}. The decision to build on existing models in \pkg{MLLib}, accessible through the \pkg{sparklyr} interface, is particularly advantageous in the context of big data processing because the underlying framework is specifically designed to efficiently handle large data sets by distributing execution over multiple cores/machines and seamlessly exchanging information between RAM and the hard drive in order to keep execution within given memory limits. This means that the models provided by \pkg{MLLib} are built to process large data sets in a scalable and computationally effective manner, which would be challenging or impossible with traditional single-machine imputation methods. Using these optimized implementations, \pkg{bigMICE} can perform imputation tasks on massive data sets, allowing the analysis of data sets that were previously too large for standard statistical software or required specialized computational facilities.

Across all these imputation functions, a consistent workflow is followed:
\begin{itemize}
    \item A temporary sequential ID is added to the input \pkg{Spark} DataFrame to preserve the original row order. The ordering of the data is important to preserve as the missingness patterns can have an influence on the quality of imputation.
    \item The DataFrame is divided into two subsets: one containing complete observations for the target variable and another containing missing values. The observed data will be used to fit the regression or classification model, which will then be used to predict the missing values.
    \item The model formula is generated, specifying the target variable as predicted by the feature variables. This is an essential parameter for the regression models in \pkg{sparklyr}.
    \item The regression model is trained using the cases with the observed values in the variable of interest.
    \item This trained model is then used to predict the missing values in the variable of interest.
    \item The missing observations in the variable of interest are replaced by data generated from marginal distributions.
    \item Finally, the complete and imputed subsets are combined, reordered based on the temporary ID, and the ID column is removed before the resulting DataFrame is returned.
\end{itemize}

The imputation functions implemented in \pkg{bigMICE} are presented in Table \ref{tab:imputation_functions}.

\begin{table}[htbp]
\centering
\caption{Imputation Functions in \pkg{bigMICE}}
\label{tab:imputation_functions}
\begin{tabularx}{\textwidth}{|>{\raggedright\arraybackslash}X|>{\raggedright\arraybackslash}p{2.2cm}|>{\raggedright\arraybackslash}p{3.4cm}|>{\raggedright\arraybackslash}p{2cm}|}
\hline
\textbf{Function Name} & \textbf{Description} & \textbf{Underlying sparklyr Function} & \textbf{Target Variable Type} \\
\hline
\texttt{impute\_with\_linear\_regression} & Linear Regression & \texttt{ml\_linear\_ regression} & Continuous \\
\hline
\texttt{impute\_with\_logistic\_regression} & Logistic Regression & \texttt{ml\_logistic\_ regression} & Binary categorical \\
\hline
\texttt{impute\_with\_mult\_logistic\_ regression} & Multinomial Logistic Regression & \texttt{ml\_logistic\_ regression} & Categorical (>2 levels) \\
\hline
\texttt{impute\_with\_random\_forest\_ regressor} & Random Forest Regressor & \texttt{ml\_random\_ forest\_regressor} & Continuous \\
\hline
\texttt{impute\_with\_random\_forest\_ classifier} & Random Forest Classifier & \texttt{ml\_random\_ forest\_classifier} & Categorical \\
\hline
\end{tabularx}
\end{table}

\subsection{Generating from marginal distribution}

\subsubsection{Continuous variables}

Root mean squared error (RMSE) is calculated first from the model residuals of the predictions on the observed values: $$\hat\sigma_{res} = \sqrt{\frac{1}{n_{obs}}\sum_{i=1}^{n_{obs}}(y_{i}^{obs} - \hat{y}_i^{obs})^2},$$ where $n_{obs}$ is the number of observed values, $y_{i}^{obs}$ represents the observed values and $\hat{y}_i^{obs}$ the corresponding predictive means from the current model. To generate from the marginal distribution corresponding to the continuous variables, random noise $\epsilon_i \sim \mathcal{N}(0, \hat\sigma_{res}^2)$ was then added to each imputation prediction: $y_{imp,i} = \hat{y}_i + \epsilon_i$. These values were used to impute the missing values. This approach follows the parametric bootstrapping idea \cite{Efron_boostrap}. This can be seen as an appropriate alternative for drawing from posterior predictive distribution for non-Bayesian models, such as Random Forests.

\subsubsection{Categorical Variables}

Missing categorical variables are imputed using a probabilistic sampling approach that preserves classification uncertainty. An appropriate classification algorithm is trained on complete cases to predict the probabilities of class membership $P(Y = c_k | X)$ for each category $c_k, k = 1, \ldots, K$. Rather than deterministically assigning the most probable class, we implemented probabilistic imputation following the inverse CDF method \cite{Devroye86}. For each observation $i$ with missing values, the model generates a probability vector $\mathbf{p}_i = [p_{i1}, p_{i2}, \ldots, p_{iK}]$ where $\sum_{k=1}^{K} p_{ik} = 1$. The cumulative probabilities are calculated as $F_{ik} = \sum_{j=1}^{k} p_{ij}$, creating a cumulative distribution function for each observation. Then a uniform random variable $U_i \sim \mathcal{U}(0,1)$ is generated, and the imputed class is determined by inverse transform sampling: $\hat{Y}_i = c_k$ if $F_{i,k-1} < U_i \leq F_{ik}$, where $F_{i,0} = 0$ by convention. 

\subsection{The bigMICE sampler}

The \texttt{sampler.spark} function is the central component to perform iterative multiple imputation within \pkg{bigMICE} as it incorporates all the aforementioned functions. It takes an initial imputed \pkg{Spark} DataFrame and iteratively imputes missing values for each variable over a specified number of cycles. For each variable, it determines the relevant predictor variables (all the other variables by default or using a user-provided predictor matrix) and the appropriate imputation method based on the variable types provided by the user. The function then calls the corresponding imputer function (\texttt{impute\_with\_...}) to perform the imputation for that specific variable in the current iteration. This process repeats for all variables and across the defined iterations, following the MICE algorithm, ultimately returning a fully imputed \pkg{Spark} DataFrame.

\subsection{The main Function and downstream analyses}

The \texttt{mice.spark()} function generates $m$ complete data sets by iteratively imputing missing values through chained equations, where each variable with missing data is modeled conditionally on all other variables. For each imputation $i = 1, \ldots, m$, the function initializes missing values using \texttt{init\_with\_random\_samples()}, performs iterative imputation via \texttt{sampler.spark()} with a maximum of \texttt{maxit} iterations, and fits a user-specified analysis model to obtain parameter estimates $\hat{\theta}_i$.

To avoid keeping each imputed data set in memory, the function performs the given downstream analyses directly after each imputation cycle and saves only the parameters of interest for subsequent pooling. Currently, the downstream analysis is limited to logistic regression modeling using the \newline\texttt{sparklyr::ml\_logistic\_regression} function with a user-provided formula. This memory-efficient approach avoids the significant storage requirements of saving complete imputed data sets, particularly beneficial for large data sets and high numbers of imputations.

Following Rubin's rules for multiple imputation inference, the function pools the $m$ parameter estimates to compute the overall estimate. Suppose that $\hat{Q}_l$ is the estimate of the $l^{th}$ repeated imputation, then the combined estimate is equal to $\bar{Q} = \frac{1}{m}\sum_{l=1}^{m}\hat{Q}_l$. The within-imputation variance is defined as $\bar{U} = \frac{1}{m}\sum_{l=1}^{m}\bar{U}_l$ where the term $\bar{U}_l$ is the variance-covariance matrix of Q obtained for the $l^{th}$ imputation. The between-imputation variance $B = \frac{1}{m-1}\sum_{l=1}^{m}(\hat{Q}_l - \bar{Q})(\hat{Q}_l - \bar{Q})^T$, and the total variance $T = \bar{U} + B + B/m$. This approach properly accounts for both the uncertainty within each imputation and the additional uncertainty introduced by the imputation process itself, enabling valid statistical inference in the presence of missing data while leveraging \pkg{Spark}'s distributed computing capabilities for scalability to large data sets. The function returns a comprehensive list containing Rubin's statistics \cite{Buuren_2018_book}, per-imputation statistics, and the saved model parameters.

\subsection{Accessing the imputed data}

In the context of big data, handling and storing multiple imputed data sets presents significant memory challenges. The preferred workflow when using the \texttt{mice.spark} function is to return only the final analysis results, thus avoiding the overhead of managing large imputed data sets in memory. However, there are situations where accessing the imputed data itself may be of interest, for instance, to assess the quality of the imputations or to conduct custom downstream analyses.

To support such use cases, the package provides the \texttt{mice.spark.plus} function, which allows retrieval of imputed data sets. The imputed data sets are stored in memory as \pkg{Spark} DataFrames and returned as part of the function output, enabling users to directly inspect and analyze the imputations. This approach can become memory-intensive for large data sets, potentially limiting scalability.

\subsection{Additional parameters}

The \texttt{mice.spark} function offers several optional parameters that allow more control over the imputation procedure and the usage of memory. A complete description of all parameters, including their purpose and usage, is provided in Appendix~\ref{sec:parameter-reference}.

\section{Experiments}

\subsection{Data: Swedish National Diabetes Registry}

The National Diabetes Registry (NDR) was created in 1996 by the Swedish Society for Diabetology. An objective of NDR is that all diabetic patients in Sweden should ideally be reported annually, based on registered annual data from actual patient visits to primary health care or at hospital outpatient clinics for departments of medicine. Demographic data, duration of diabetes, treatment modalities, as well as various risk factors and diabetes complications are reported \cite{NDR_2003}. NDR is probably among the largest national diabetes registers in the world \cite{NDR_2003}.
This data set covers the majority of patients with type 2 diabetes in Sweden and contains multiple data points per individual over time \cite{NDR_2003}. It contains approximately 14.6 million rows and 58 variables of different kind, each row representing a patient record at a specific time point. The proportion of missingness for each variable in this data set varies between 0 and 99.96\%, see supplementary Table \ref{tab:ndr_variable_metadata}.

\subsection{Dependence of runtime and memory usage on sample sizes}

To evaluate how the computational runtime and memory usage of \pkg{bigMICE} scales with increasing data set sizes, a series of experiments were performed using the National Diabetes Register data set. A subset of 10 variables was selected to include different types of variables (binary, integer, float, ordinal, and nominal), as well as varying proportions of missing values in the variables, from 0\% to 65.8\%. Linear regression was used to impute integer and float variables, logistic regression was used for binary variables, multinomial logistic regression was used for nominal variables, and the random forest classifier method was used for ordinal variables. For each variable's conditional model, all other variables were used as predictors and the default parameter of the respective \pkg{MLLib} functions were used.


Data sets were generated by random sampling without replacement from the original data set, with logarithmically increasing sizes from 1000 rows to the full data set of 14.6 million rows. To run our experiments, we have used a computational server with 512 Gb of RAM and a 32 core processor with 64 threads in total. For \pkg{bigMICE}, execution was restricted to 8 threads and 16 Gb of RAM using the \pkg{Spark} session parameters, while \pkg{mice}, the baseline implementation, had access to all threads and full memory. For each sample size, the MICE imputation procedure was applied using the \pkg{bigMICE} implementation and the \pkg{mice} implementation for baseline comparison.

The number of imputations was set to m=1, with 5 iterations performed per imputation (default setting), and the procedure was repeated five times for each sample size to account for variability in system performance and runtime stability. The same imputation models and parameters were used in the \pkg{bigMICE} method and the baseline, with the exception of ordinal variables, where we used random forest in \pkg{bigMICE}, and random forest or polytomous regression in \pkg{mice}, due to reasons explained later in this section. The details of variable types and models used for the imputation can be found in Table \ref{tab:exp1_var_explain}.

\begin{table}[h]
    \centering
    \begin{tabular}{|l|c|p{4cm}|p{4.5cm}|}
        \hline
        \textbf{Variable Name} & \textbf{Data Type} & \textbf{Baseline Models (\texttt{function})} & \textbf{bigMICE Models (\texttt{function})} \\
        \hline
        alder & Integer & Bayesian linear regression (\texttt{norm}) & Linear regression (\texttt{ml\_linear\_regression}) \\
        \hline
        bmi & Float & Bayesian linear regression (\texttt{norm}) & Linear regression (\texttt{ml\_linear\_regression}) \\
        \hline
        kolesterol & Float & Bayesian linear regression (\texttt{norm}) & Linear regression (\texttt{ml\_linear\_regression}) \\
        \hline
        sex & Nominal & Multinomial Logistic regression (\texttt{polyreg}) & Multinomial Logistic regression (\texttt{ml\_logistic\_regression}) \\
        \hline
        klin\_diab\_typ & Nominal & Multinomial Logistic regression (\texttt{polyreg}) & Multinomial Logistic regression (\texttt{ml\_logistic\_regression}) \\
        \hline
        hba1c & Integer & Bayesian linear regression (\texttt{norm}) & Linear regression (\texttt{ml\_logistic\_regression}) \\
        \hline
        stroke & Binary & Logistic regression (\texttt{logreg}) & Logistic regression (\texttt{ml\_logistic\_regression}) \\
        \hline
        fysisk\_aktivitet & Ordinal & Random forest (\texttt{rf}) or Polytomous regression (\texttt{polr}) & Random Forest (\texttt{ml\_random\_forest}) \\
        \hline
        rokare & Nominal & Multinomial Logistic regression (\texttt{polyreg}) & Multinomial Logistic regression (\texttt{ml\_logistic\_regression}) \\
        \hline
        hypoglykemi & Ordinal & Random forest (\texttt{rf}) or Polytomous regression (\texttt{polr}) & Random Forest (\texttt{ml\_random\_forest}) \\
        \hline
    \end{tabular}
    \caption{Variables used in the memory usage and runtime measurement experiment, and the associated models used for the baseline and \pkg{Spark} methods. }
    \label{tab:exp1_var_explain}
\end{table}

Memory consumption is measured in different ways for the baseline and the \pkg{bigMICE} method, as their computation happens differently. For the baseline, the memory usage of the \proglang{R} process that runs the code is measured during the execution of the \texttt{mice()} function that produces the imputations. For the \pkg{bigMICE} procedure, this approach is not possible, as the computation is performed by the Java Virtual Machine (JVM), inside which the \pkg{Spark} session is running, and thus it is not possible to capture the memory consumption by monitoring the \proglang{R} process. Instead, we collect metrics by enabling metrics collection in the \pkg{Spark} connection and retrieving the metrics from the \pkg{Spark} UI after the execution of the procedure. More specifically, the peak JVM heap usage (\texttt{peakMemoryMetrics.JVMHeapMemory}) is used to monitor the maximum memory consumption of the procedure during execution. The detailed implementation of the code used to monitor the memory is presented in the appendix \ref{sec:memory_code}. 

For runtime measurements, only the computational time of the imputation procedure is measured. All other computations, such as loading the libraries, loading the data into \proglang{R} or \pkg{Spark}, or starting the \pkg{Spark} connection, are excluded from the runtime measurements.

\begin{figure}
    \centering
    \includegraphics[width=1\linewidth]{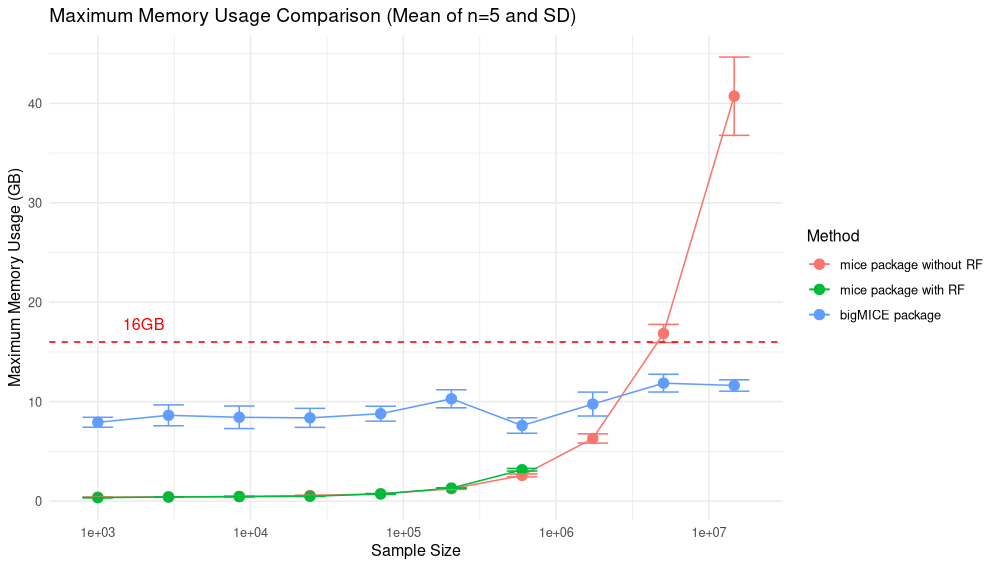}
    \caption{Memory usage comparison between the baseline package \pkg{mice} and package \pkg{bigMICE}}
    \label{fig:mem}
\end{figure}
\begin{figure}
    \centering
    \includegraphics[width=1\linewidth]{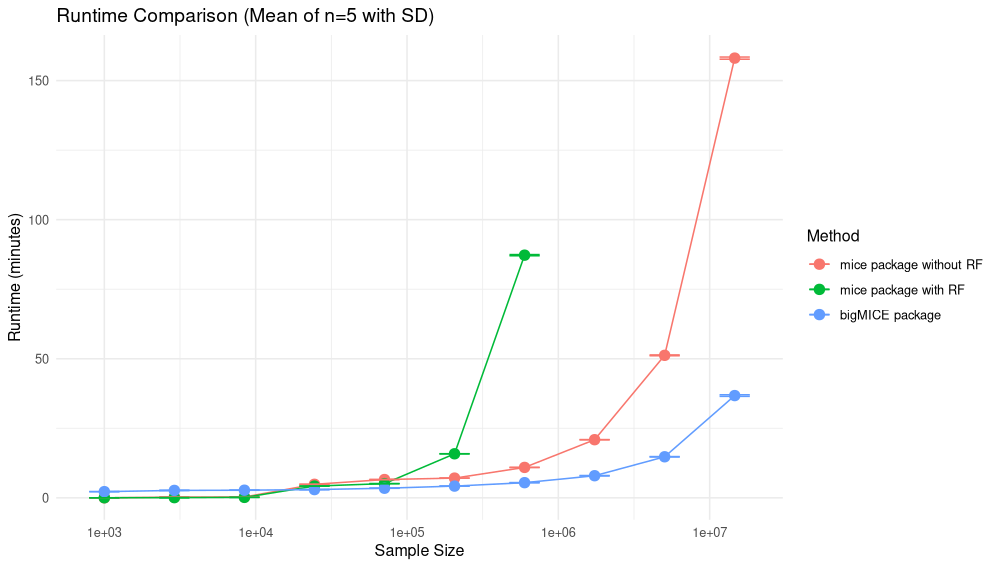}
    \caption{Runtime comparison between the baseline package \pkg{mice} and package \pkg{bigMICE}}
    \label{fig:run}
\end{figure}

The results of this experiment are presented in Figure \ref{fig:mem} for the memory and Figure \ref{fig:run} for the runtime. In addition, the same results are presented in Table \ref{tab:performance}. The red line in the memory plot represents the maximum memory available on a typical laptop in 2025 (16Gb).


\begin{table}[ht]
\centering
\begin{tabular}{rlll}
  \toprule
Sample Size & Method & Memory (Gb) (mean $\pm$ sd) & runtime (minutes) (mean $\pm$ sd) \\ 
  \midrule
   & mice & 0.391 $\pm$ 0.000 & 0.012 $\pm$ 0.001 \\ 
   1000 & mice (with RF) & 0.348 $\pm$ 0.015 & 0.020 $\pm$ 0.001 \\ 
     & bigMICE & 7.929 $\pm$ 0.500 & 2.219 $\pm$ 0.025 \\ 
  \hline
    & mice & 0.430 $\pm$ 0.000 & 0.260 $\pm$ 0.001 \\ 
  2902 & mice (with RF) & 0.410 $\pm$ 0.012 & 0.061 $\pm$ 0.002 \\ 
    & bigMICE & 8.627 $\pm$ 1.050 & 2.659 $\pm$ 0.030 \\ 
  \hline
    & mice & 0.462 $\pm$ 0.008 & 0.313 $\pm$ 0.002 \\ 
  8426 & mice (with RF) & 0.455 $\pm$ 0.009 & 0.156 $\pm$ 0.002 \\ 
    & bigMICE & 8.429 $\pm$ 1.133 & 2.736 $\pm$ 0.018 \\ 
  \hline
    & mice & 0.563 $\pm$ 0.007 & 4.861 $\pm$ 0.002 \\ 
  24459 & mice (with RF) & 0.483 $\pm$ 0.007 & 4.213 $\pm$ 0.004 \\ 
    & bigMICE & 8.371 $\pm$ 0.954 & 2.957 $\pm$ 0.017 \\ 
  \hline
    & mice & 0.732 $\pm$ 0.011 & 6.569 $\pm$ 0.007 \\ 
  70999 & mice (with RF) & 0.723 $\pm$ 0.036 & 5.105 $\pm$ 0.022 \\ 
    & bigMICE & 8.789 $\pm$ 0.751 & 3.438 $\pm$ 0.017 \\ 
  \hline
    & mice & 1.242 $\pm$ 0.065 & 7.084 $\pm$ 0.019 \\ 
  206096 & mice (with RF) & 1.307 $\pm$ 0.017 & 15.829 $\pm$ 0.033 \\ 
    & bigMICE & 10.290 $\pm$ 0.908 & 4.216 $\pm$ 0.043 \\ 
  \hline
    & mice & 2.579 $\pm$ 0.145 & 10.930 $\pm$ 0.025 \\ 
  598253 & mice (with RF) & 3.154 $\pm$ 0.130 & 87.238 $\pm$ 0.219 \\ 
    & bigMICE & 7.601 $\pm$ 0.775 & 5.446 $\pm$ 0.036 \\ 
  \hline
    & mice & 6.299 $\pm$ 0.465 & 20.897 $\pm$ 0.043 \\ 
  1736597 & mice (with RF) & NaN $\pm$ NA & NaN $\pm$ NA \\ 
    & bigMICE & 9.758 $\pm$ 1.202 & 7.943 $\pm$ 0.085 \\ 
  \hline
    & mice & 16.856 $\pm$ 0.914 & 51.226 $\pm$ 0.152 \\ 
  5040959 & mice (with RF) & NaN $\pm$ NA & NaN $\pm$ NA \\ 
    & bigMICE & 11.863 $\pm$ 0.893 & 14.754 $\pm$ 0.065 \\ 
  \hline
    & mice & 40.723 $\pm$ 3.939 & 158.088 $\pm$ 0.391 \\ 
  14632799 & mice (with RF) & NaN $\pm$ NA & NaN $\pm$ NA \\ 
    & bigMICE & 11.629 $\pm$ 0.570 & 36.754 $\pm$ 0.285 \\ 
   \bottomrule
\end{tabular}
\caption{Performance Comparison} 
\label{tab:performance}
\end{table}

The baseline measurements were performed twice with a slightly different configuration. In one of the configurations, we used random forest to imputed ordinal values, but the runtime grew very rapidly, long before we could reach a point where it would potentially use too much memory, so we had to stop the experiment. Therefore, we ran another baseline configuration, where the random forest model was replaced by polytomous regression, which allowed us to demonstrate that the memory needed for the computation grew rapidly with the increasing sample size. 

Figure \ref{fig:mem} shows that the developed method demonstrate constant memory usage in all sample sizes, not exceeding the allocated 16Gb of memory, and improved runtime over the baseline methods in the larger sample sizes. This demonstrates the ability of \pkg{bigMICE} to impute missing values on very large data sets while maintaining control over the memory used.

\subsection{Dependence of the runtime on the number of variables included}

To evaluate how the runtime of the \pkg{bigMICE} method scales with increasing number of variables included, a series of experiments were performed using the National Diabetes Register data set. 

A random subset of one thousand rows was selected. Linear regression was used to impute integer and float variables, logistic regression was used for binary variables, multinomial logistic regression was used for nominal variables, and the random forest classifier method was used for ordinal variables.
We performed seven measurements for the numbers of variables ranged from 5 to 49 in increments of 5, to assess how the runtime changes. To mitigate the effect of a variable's missingness proportion on the runtime, the variables were added in a different order during each measurement. Two runs followed a determined order: increasing missingness proportion and decreasing missingness proportion, while the remaining 5 runs were performed by randomly permuting the order of variables.

\begin{figure}
    \centering
    \includegraphics[width=0.8\linewidth]{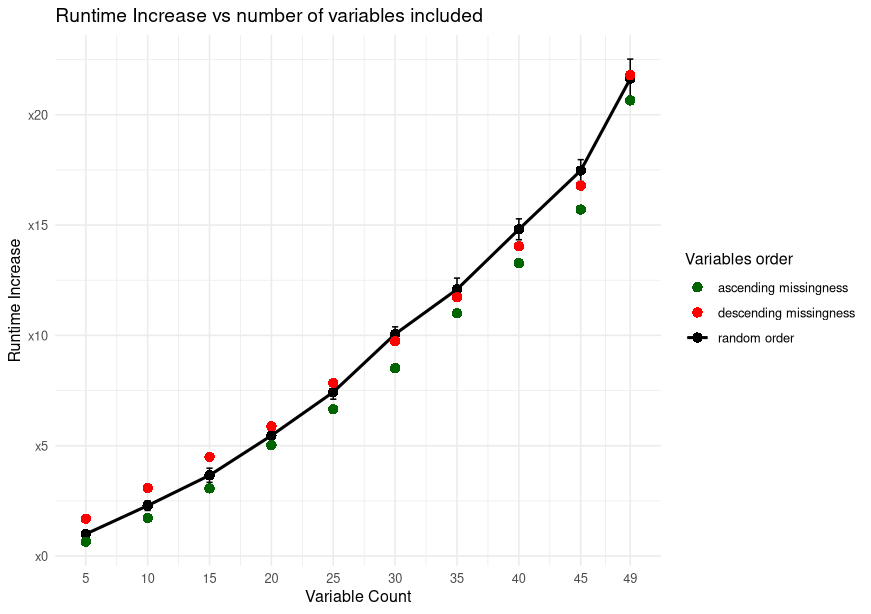}
    \caption{Runtime comparison for subsets with different numbers of variables.}
    \label{fig:variable_exp}
\end{figure}

The results are presented in Figure \ref{fig:variable_exp} and in Table \ref{tab:runtime_summary}. It can be observed in Figure \ref{fig:variable_exp} that the runtime grows at a rate that is slightly greater than linear. In addition, the missingness proportion of the variable being imputed appears to have a minor but noticeable effect on runtime. This is evident in the measurements with fewer variables, where the descending missingness order (imputing variables with the highest missingness proportion first) results in a runtime slightly higher than the average.

\begin{table}[ht]
\centering
\begin{tabular}{rrrr}
  \toprule
Variable Count & Mean Runtime (sec) & Std Dev (sec) & Runtime Increase \\ 
  \midrule
5 & 32.98 & 5.00 & 1.00x \\ 
  10 & 75.43 & 7.20 & 2.29x \\ 
  15 & 120.72 & 10.63 & 3.66x \\ 
  20 & 180.15 & 9.91 & 5.46x \\ 
  25 & 245.06 & 10.62 & 7.43x \\ 
  30 & 331.78 & 10.82 & 10.06x \\ 
  35 & 398.64 & 16.78 & 12.09x \\ 
  40 & 488.36 & 15.62 & 14.81x \\ 
  45 & 576.16 & 16.30 & 17.47x \\ 
  49 & 713.34 & 29.35 & 21.63x \\ 
   \bottomrule
\end{tabular}
\caption{Runtime summary: mean, standard deviation, and relative increase compared to the runtime with 5 variables } 
\label{tab:runtime_summary}
\end{table}

\subsection{Quality of imputation vs missingness proportion and sample size}

To assess the quality of the imputations produced by the \pkg{bigMICE} method, two complementary experiments were designed: (1) varying the sample size while keeping an approximately constant missingness proportion and (2) varying the missingness proportion with a constant sample size.

The continuous variable \texttt{GFR} from the National Diabetes Register data set was selected as the target for these experiments. This variable originally had a low proportion of missing values, allowing for controlled artificial masking of entries. This setup enables a precise evaluation of the imputation quality by comparing the imputed values to their true known counterparts.
The quality of imputation was quantified using the Root Mean Squared Error (RMSE) for the \texttt{GFR} variable of the imputed values compared to the true values.
The relevant subsets for each experimental condition were then generated as follows. 
For the varying sample size experiment, data sets were generated with six different sample sizes: 1,000, 5,000, 9,503, 90,303, 858,129, and 8,154,612 rows. These sizes were chosen to cover a broad spectrum of practical scenarios, ranging from small-scale analyses to very large data sets. The intermediate values were selected to capture both moderate and large scales in a roughly logarithmic progression, while the largest sample size corresponds to all observed \texttt{GFR} rows in the dataset.
The proportion of missing values was kept at $\sim$50\% in all sample sizes. 
For the experiment on varying missingness proportion, a fixed sample size of 1 million rows was used. The proportion of artificially removed entries was set to 1\%, 10\%, 30\%, 50\%, 70\%, 90\%, 99\%, and 99.9\% to cover a wide range of missingness scenarios, from very low to extreme levels. Extreme missingness proportions were included because the dataset is very large, allowing meaningful evaluation even when a substantial fraction of values is removed.

In both experiments, missing values were introduced by randomly masking a necessary number of values in the variable of interest. This simulates a scenario where the variable is missing completely at random (MCAR). For each condition, $m=5$ imputations were generated, each imputation performed over $5$ iterations. RMSE was calculated for each imputation.

We used random forest to impute the GFR variable. The rest of the imputation models used for the selected variable are presented in supplementary Table \ref{tab:variables_models_quality}.

\begin{table}[ht]
\centering
\caption{Overview of variables, types, and imputation models used}
\begin{tabular}{lll}
\hline
\textbf{Variable} & \textbf{Type} & \textbf{Imputation Model} \\
\hline
alder          & Integer           & Linear regression \\
sex            & Nominal           & Multi Logistic regression  \\
hba1c          & Integer           & Linear regression  \\
klin\_diab\_typ & Nominal           & Random Forest classifier \\
debutar        & Integer           & Linear regression  \\
diab\_beh      & Nominal           & Random Forest classifier \\
blodtrycksmed  & Binary            & Logistic regression  \\
lipidmed       & Binary            & Logistic regression  \\
asa            & Binary            & Logistic regression  \\
langd          & Integer           & Linear regression  \\
systoliskt     & Integer           & Linear regression  \\
diastoliskt    & Integer           & Linear regression  \\
kreatinin      & Integer           & Linear regression  \\
\textbf{GFR}   & \textbf{Float}    & \textbf{Random Forest regression } \\
retinopati     & Binary            & Logistic regression  \\
bmi            & Float             & Linear regression  \\
kolesterol     & Float             & Linear regression  \\
ldl            & Float             & Linear regression  \\
hdl            & Float             & Linear regression  \\
albuminuria    & Nominal           & Random Forest classifier \\
\hline
\end{tabular}
\label{tab:variables_models_quality}
\end{table}

The results of the experiment with varying sample sizes are presented in supplementary Table \ref{tab:sample_size_res}. The results of the experiment with varying missingness proportions are presented in supplementary Table \ref{tab:miss_res}. Supplementary Table \ref{tab:sample_size_res} illustrates that the imputation error has an overall decreasing trend with increasing sample size, which is an expected model behavior. Supplementary Table \ref{tab:miss_res} reveals that for such large data set with 1 million rows, a notable decrease of imputation accuracy starts only at 99.9\% missingness proportion, while for all other considered missingness proportion the imputation accuracy is quite similar.

\begin{table}[ht]
\centering
\begin{tabular}{rrrr}
  \toprule
Sample Size & Mean RMSE & Standard Deviation \\ 
  \midrule
1000 & 1.70038 & 0.003850 \\ 
5000 & 1.68944 & 0.002221 \\ 
9503 & 1.62626 & 0.001940 \\ 
90303 & 1.63172 & 0.001524 \\ 
858129 & 1.61430 & 0.002354 \\ 
8154612 & 1.62250 & 0.002431 \\ 
   \bottomrule
\end{tabular}
\caption{Effect of the sample size on quality of imputation of the GFR variable when missing proportion is 50\% (m=5 imputations)} 
\label{tab:sample_size_res}
\end{table}

\begin{table}[ht]
\centering
\begin{tabular}{lrrr}
  \toprule
Missingness Proportion (\%) & Mean RMSE & Standard Deviation \\ 
  \midrule
10 & 1.61979 & 0.000610  \\ 
30 & 1.61666 & 0.000276  \\ 
50 & 1.61509 & 0.000254  \\ 
70 & 1.61634 & 0.002322  \\ 
90 & 1.61088 & 0.000393  \\ 
99 & 1.62880 & 0.001507  \\ 
99.9 & 1.70810 & 0.003224  \\ 
   \bottomrule
\end{tabular}
\caption{Effect of the missingness proportion on quality of imputation of the GFR variable with 1 million samples (m=5 imputations)} 
\label{tab:miss_res}
\end{table}

\section{Discussion}

\subsection{Discussion of Results}

The \pkg{bigMICE} method has similar memory usage during imputation in all tested sample sizes. In contrast, the baseline memory usage increases rapidly with the sample size. This difference comes from \pkg{Spark}’s ability to dynamically exchange data between RAM and the hard drive, offloading data to the hard drive when RAM capacity is reached. In particular, measured RAM usage does not reach the full 16 GB allocation because we only track peak On Heap memory, the space dedicated to computation, while \pkg{Spark} reserves the remaining allocated memory for auxiliary tasks, such as metadata storage, or other system operations \cite{karau2017spark}.

In addition to the memory efficiency, \pkg{bigMICE} also outperforms the baseline in terms of the runtime when imputing large data sets. This improvement is attributed to the distributed execution of base learners in \pkg{Spark} and the use of machine learning model implementations with low computational complexities. In addition, the baseline method relies on a single-threaded computation, resulting in a runtime that grows more rapidly with the sample size. This limitation was particularly evident in the experiments using random forests, where the runtime became prohibitively long, forcing premature termination of the tests.

The runtime also depends on the number of variables being imputed. The runtime increases with the increasing number of variables, but the increment becomes slightly more pronounced when exceeding 30 variables. This effect likely stems from the increased complexity of fitting models in high-dimensional feature spaces. Additionally, runtime performance may be influenced by hard drive read/write speeds and the network latency, especially if data storage and computations occur on separate machines. Checkpointing operations play a crucial role in this context. Checkpointing enables the method to segment the execution of large computations into manageable chunks, mitigating memory constraints by periodically saving intermediate results to disk. This mechanism is essential for processing large datasets with limited memory, as it prevents out-of-memory errors and ensures stability, but at the cost of introducing I/O overhead that directly impacts runtime.

The experiment with varying sample size shows that the quality of imputation is improved by having more samples, despite having a large proportion (50\%) of missing values. This is because large data still have many observed cases, even when missingness proportion is as high as 50\%.
This finding is further emphasized in the experiment with varying proportion of missingness in a sample size of 1 million rows. These results show that the mean RMSE does not change much until missingness proportion becomes extreme, i.e. 99.9\%.

This demonstrates a big potential of \pkg{bigMICE} for performing multiple imputation on very large data sets. The abundance of observed values enables good  imputation quality, even when missingness proportion is very high. However, it is important to note that our experiments assume data is Missing Completely at Random (MCAR), which may not always reflect real-world scenarios. For example, in data sets like healthcare records, variables such as PumpPagaendeModell might be missing not at random but because patients do not use a pump. Such Missing Not at Random (MNAR) mechanisms could introduce bias if not accounted for. Nonetheless, the robustness of \pkg{bigMICE} in imputing large data sets with high missingness proportions opens new opportunities in the research that previously excluded some variables due to high missingness proportions.

\subsection{Future works}

Although the current implementation is efficient for Big Data, further development is required to achieve ease of use similar to the original package \pkg{mice}, a widely adopted tool for multiple imputation, although limited in its ability to handle very large datasets.
Key directions for future work include the following.

\begin{itemize}
\item Expanding Imputation Model Support:
The range of supported imputation models could be broadened to accommodate more use cases and data types. For example, additional \pkg{MLLib} models available through \pkg{sparklyr} have not yet been integrated into \pkg{bigMICE}. 
\item Enhancing Functionality and User Control:
To match the flexibility of \pkg{mice}, the package should incorporate features such as custom imputation sequences, convergence diagnostics, and user-defined model parameters. User feedback could help identify the most critical functionalities to implement first.
\item Optimizing \pkg{Spark}-Based Procedures:
\pkg{Spark} workflows could be optimized to reduce computational overhead and improve runtime efficiency. For example, achieving the same results with fewer or more efficient \pkg{Spark} operations might lower computational costs. 
\item Evaluating Cluster Computing Performance:
One of the primary advantages of implementing MICE with \pkg{Spark} is the potential for distributed computation. Our research used a standalone multicore machine, but evaluating performance in a cluster environment would also be valuable. However, inter-node data transfer volumes may introduce additional overhead. Benchmarking against standalone performance could clarify the trade-offs and benefits of distributed execution. Access to a cluster would be necessary to explore this further.

\end{itemize}

\section{Conclusion}

The package \pkg{bigMICE} represents a successful first step toward extending the MICE framework to Big Data scenarios. It offers explicit control over RAM consumption during the imputation process, while also achieving faster computational times on very large datasets compared to a popular MICE implementation. The method scales efficiently with increasing sample sizes, both in terms of memory usage and runtime, enabling multiple imputation for Big Data to be performed within a reasonable timeframe — even on modest computational resources such as ordinary laptops.

\section*{Data Availability}

The NDR data set can be accessed upon request to the registry organization. A prerequisite for conducting research on data in the Diabetes Registry is that there is approval from the ethics review board, approval of data disclosure, and confidentiality assessment by the central data controller in the V\"astra G\"otaland Region \cite{NDR_web}.

\section*{Acknowledgments}

\begin{leftbar}
 F.I is an associated clinical fellow of Wallenberg Center for Molecular Medicine (WCMM) and receives financial support from the Knut and Alice Wallenberg Foundation.

\end{leftbar}


\bibliography{references}


\newpage

\begin{appendix}

\section{Parameter Reference for mice.spark()}
\label{sec:parameter-reference}

This appendix provides a detailed overview of the parameters available in the \texttt{mice.spark()} function.

\paragraph{sc.} 
A Spark connection created with \texttt{sparklyr::spark\_connect()}. This provides the context in which all data processing and modeling will be executed.

\paragraph{data.} 
A Spark DataFrame containing the data set with missing values. The function operates directly on this distributed data object without requiring conversion to in-memory \texttt{R}  data frames.

\paragraph{variable\_types.} 
A named character vector specifying the type of each column in the data set (e.g., \texttt{"Continuous\_int"}, \texttt{"Binary"}, \texttt{"Nominal"}). These types guide the initialization procedure and the choice of imputation method.

\paragraph{analysis\_formula.} 
A formula describing the analysis model of interest, such as \texttt{outcome ~ age + income + education}. This formula is passed to Spark MLLib to fit the analysis model after each imputation.

\paragraph{m.} 
The number of multiple imputations to generate. Default is 5. Larger values reduce Monte Carlo error but increase computational cost.

\paragraph{method.} 
Specifies the imputation strategy used for each variable. If \texttt{NULL}, defaults are inferred from \texttt{variable\_types}. Users can override by supplying a character vector indicating, for example, regression-based, logistic, or multinomial imputation.

\paragraph{predictorMatrix.} 
An optional binary matrix defining which variables are used as predictors for imputing others. Rows correspond to variables being imputed, and columns to potential predictors. By default, all other variables are used except the variable itself.

\paragraph{formulas.} 
A list of Spark-compatible formulas specifying custom imputation models for individual variables. Overrides the predictor structure implied by \texttt{predictorMatrix}.

\paragraph{modeltype.} 
Specifies the type of statistical model applied for imputation, such as linear regression, logistic regression, or multinomial regression. If \texttt{NULL}, inferred from \texttt{variable\_types}.

\paragraph{maxit.} 
The maximum number of iterations of chained equations to perform per imputation. Default is 5. Larger values allow more thorough convergence but require more computation.

\paragraph{printFlag.} 
Logical flag controlling whether diagnostic and progress messages are printed during execution. Useful for debugging or monitoring long computations.

\paragraph{seed.} 
An integer random seed used for reproducibility of imputations. If \texttt{NA}, a random seed is chosen.

\paragraph{imp\_init.} 
An optional Spark DataFrame containing the data after an initial round of imputation (e.g., using mean, median, or random draws). If not supplied, initialization is handled internally.

\paragraph{checkpointing.} 
Logical flag indicating whether Spark checkpointing should be enabled. Default is \texttt{TRUE}. Recommended for preventing lineage growth and potential stack overflow errors in Spark.

\paragraph{checkpoint\_frequency.} 
Controls how often checkpointing is triggered when \texttt{checkpointing = TRUE}. Default is 10, meaning checkpointing occurs after every 10 variables processed. Lower values reduce lineage growth at the cost of increased disk I/O.

\section{Additional Tables}

\begin{longtable}{lrl}
\caption{NDR registry variable metadata summary.}
\label{tab:ndr_variable_metadata}
\\
\toprule
\textbf{Variable Name} & \textbf{Missing (\%)} & \textbf{Data Type} \\
\midrule
\endfirsthead

\multicolumn{3}{c}{\tablename\ \thetable\ -- continued from previous page} \\
\toprule
\textbf{Variable Name} & \textbf{Missing (\%)} & \textbf{Data Type} \\
\midrule
\endhead

\bottomrule
\multicolumn{3}{r}{Continued on next page} \\
\endfoot

\bottomrule
\endlastfoot

alder & 0.00 & Integer \\
LopNr & 0.00 & Numerical (code) \\
regdat & 0.00 & smalldatetime \\
SenPNr & 0.00 & Numerical (code) \\
sex & 0.00 & Nominal \\
klin\_diab\_typ & 0.09 & Nominal \\
diab\_beh / treatment & 3.19 & Nominal \\
blodtrycksmed & 3.57 & Binary \\
debutar & 3.68 & Integer \\
lipidmed & 3.95 & Binary \\
asa & 5.18 & Binary \\
klin\_diab\_typ\_reg & 9.41 & Numerical (code) \\
langd & 9.63 & Integer \\
ischemisk\_hjsjukdom & 11.87 & Binary \\
stroke & 12.97 & Binary \\
fotundersokning & 18.43 & Numerical (code) \\
fotrisk & 24.01 & Nominal \\
ogon\_datum & 25.27 & smalldatetime \\
fot\_datum & 25.67 & smalldatetime \\
retinopati & 31.20 & Binary \\
hba1c & 31.62 & Integer \\
systoliskt & 34.64 & Integer \\
diastoliskt & 34.78 & Integer \\
debutar\_reg & 38.90 & Numerical (code) \\
kreatinin & 44.25 & Integer \\
GFR & 44.27 & Float \\
rokare & 45.05 & Nominal \\
vikt & 45.27 & Float \\
bmi & 47.21 & Float \\
rokvanor & 54.94 & Ordinal \\
kolesterol & 55.63 & Float \\
ldl & 56.25 & Float \\
fysisk\_aktivitet & 56.58 & Ordinal \\
albuminuria & 57.89 & Nominal \\
hdl & 58.11 & Float \\
triglycerider & 62.31 & Float \\
hypoglykemi & 65.77 & Ordinal \\
ins\_metod & 66.98 & Nominal \\
synnedsattning & 72.92 & Binary \\
dodsdatum & 78.91 & smalldatetime \\
snuffingHabit & 81.63 & Ordinal \\
CGM & 87.84 & Binary \\
uAlbCreatinine & 89.72 & Float \\
eyeTreated & 90.38 & Binary \\
retinopathyDiagnosis & 90.72 & Nominal \\
isRemote & 93.07 & Binary \\
slutatrokaar & 95.01 & Integer \\
insulin & 95.14 & Binary \\
CGMTyp & 95.92 & Nominal \\
shareGlucoseLast2W & 98.13 & Integer \\
shareGlucoseRange & 98.15 & Integer \\
carbohydrate & 98.44 & Binary \\
PumpPagaendeModell & 98.55 & Nominal \\
meanGlukosesLast2W & 98.67 & Float \\
sdCGMLast2W & 99.00 & Float \\
PumpIndikation & 99.34 & Nominal \\
snuffingEndYear & 99.56 & Integer \\
PumpAvslutOrsak & 99.96 & Nominal \\
\end{longtable}

\section{Additional code examples}

\subsection{Memory Monitoring}
\label{sec:memory_code}

\subsubsection[Monitoring the memory of an R function execution]{Monitoring the memory of an \proglang{R} function execution}

The memory monitoring of an \proglang{R} function execution is done by monitoring the memory usage of the parent process. The function \texttt{monitor\_memory} monitors the memory memory of the \proglang{R} script by creating a parallel process that tracks the memory of its parent process during execution. The following example show how to use the \texttt{monitor\_memory} function to monitor the memory and runtime of a simple procedure. 

\begin{CodeChunk}
    \begin{CodeInput}
R> mice_procedure <- function(){
R>   library(mice)
R>   sim_data <- data.frame(
R>     ID = 1:10000,
R>     Age = sample(18:80, 10000, replace = TRUE),
R>     Income = round(rnorm(10000, mean = 50000, sd = 15000), 0),
R>     Score = runif(10000, min = 0, max = 100),
R>     Category = sample(c("A", "B", "C"), 10000, replace = TRUE)
R>   )
R>   amputed_data <- ampute(sim_data, prop = 0.2)$amp
R>   imputed_data <- mice(amputed_data, m = 5, method = 'pmm', seed = 123)
R>   
R>   return(imputed_data)
R> }

R> res <- monitor_memory(
R>   func = mice_procedure
R> )

R> max_mem_during <- max(res$memory_data$memory_r_mb[res$memory_data$phase=="during"])

R> print(max_mem_during)
[1] 486.5859

R> print(res$run_time)
[1] 1.765408

    \end{CodeInput}
\end{CodeChunk}

\subsubsection{Monitoring memory of a Spark application}

Monitoring a \pkg{Spark} application memory usage is a bit more straightforward as \pkg{Spark} UI already provides many metrics measurements during runtime execution. To collect those metrics, we simply make a GET request to the spark UI, and extract the metrics of interest from the return JSON file. The following code shows how to enable metrics collection in a sparck session, then extract those metrics after a spark job is complete.

\begin{CodeChunk}
    \begin{CodeInput}
R> get_spark_ui_port <- function(sc) {
R>   sc_context <- spark_context(sc)
R>   ui_url_option <- invoke(sc_context, "uiWebUrl")
R>   if (is.null(ui_url_option)) {
R>     warning("Spark UI URL not available.")
R>     return(NA_integer_)
R>   }
R>   ui_url <- invoke(ui_url_option, "get")
R>   port_str <- str_extract(ui_url, "(?<=:)\\d+")
R>   return(as.integer(port_str)) # Convert to integer and return
R> }

R> monitor_spark_memory <- function(sc, physical_limit_gb = 16) {
R>   app_id <- spark_context_config(sc)$spark.app.id
R>   ui_port <- get_spark_ui_port(sc) # usually 4040, but can be 4041, 4042,...
R>   if(is.null(ui_port)) message("Could not find ui port")
R>   # Try to connect to Spark UI
R>   endpoint <- paste0("http://localhost:", ui_port, "/api/v1/applications/", app_id,
        "/executors")
R>   response <- GET(endpoint, timeout(5))
R>   if (status_code(response) == 200) {
R>     data <- content(response, "text") %>% fromJSON(flatten = TRUE)
R>     summary_tbl <- data.frame(
R>       executor_id = data$id,
R>       total_cores = data$totalCores,
R>       max_memory_gb = data$maxMemory / (1024^3),
R>       memory_used_gb = data$memoryUsed / (1024^3),
R>       peak_jvm_heap_memory = data$peakMemoryMetrics.JVMHeapMemory / (1024^3),
R>       total_gc_time_seconds = data$peakMemoryMetrics.TotalGCTime / 1000,
R>       disk_used_gb = data$diskUsed / (1024^3),
R>       stringsAsFactors = FALSE
R>     )
R>     print("Memory Statistics Summary:")
R>     print(summary_tbl)
R>     return(list(
R>       port_used = ui_port,
R>       app_id = app_id,
R>       raw_data = data,
R>       summary = summary_tbl
R>     ))
R>   }
R>   message("Unable to connect to Spark UI and retrieve metrics.")
R>   return(NULL)
R> }
    \end{CodeInput}
\end{CodeChunk}

Then we can use the above functions to retrieve the metrics of interest at the end of a Spark job:

\begin{CodeChunk}
    \begin{CodeInput}
R> conf <- spark_config()
R> # Basic parameters
R> conf$`sparklyr.shell.driver-memory` <- "16g"
R> conf$`sparklyr.shell.executor-memory` <- "16g"
R> conf$spark.memory.fraction <- 0.9
R> conf$`sparklyr.cores.local` <- 8
R> conf$`spark.local.dir` <- "/local/data/spark_tmp/"

R> # Enable executor metrics
R> conf$spark.metrics.executorMetricsSource.enabled <- "true"
R> conf$spark.eventLog.enabled <- "true"
R> conf$spark.eventLog.dir <- "./"
R> conf$spark.eventLog.logStageExecutorMetrics <- "true"

R> sc <- spark_connect(master = "local", 
R>                     config = conf,
R>                     spark_home = "/home/user/spark/spark-4.0.0-bin-hadoop3")

R> spark_set_checkpoint_dir(sc, "/local/data/spark_tmp/" )

R> sdf <- spark_read_csv(sc, "path/to/data", infer_schema = TRUE, null_value = "NA")

R> variable_types <- c(alder = "Continuous_int", 
R>                     bmi = "Continuous_float", 
R>                     stroke = "Binary")

R> start_time <- Sys.time()

R> res <- bigMICE::mice.spark(data = sdf,
R>                             sc = sc,
R>                             variable_types = variable_types,
R>                             analysis_formula = as.formula("stroke ~ alder + bmi"),
R>                             predictorMatrix = NULL,
R>                             m = 2,
R>                             maxit = 5)

R> end_time <- Sys.time()
R> # Collect memory data and close connection
R> spark_mem <- monitor_spark_memory(sc = sc)

R> spark_disconnect(sc)

R> runtime <- as.numeric(difftime(end_time, start_time, units="secs"))
R> max_spark_memory <- spark_mem$summary$peak_jvm_heap_memory * 1000 # Gb -> Mb

    \end{CodeInput}
\end{CodeChunk}
\end{appendix}


\end{document}